\documentclass{eptcs}
\usepackage{amssymb}
\usepackage[latin1]{inputenc}
\usepackage{amsmath}
\usepackage{amsthm}
\usepackage{graphicx}
\usepackage{pst-tree,pst-node}
\usepackage{mathpartir}

\newtheorem{theorem}{Theorem}[section]
\newtheorem{definition}[theorem]{Definition}
\newtheorem{example}{Example}

\def \tuple#1{\langle #1 \rangle}

\def\defemb#1#2{\expandafter\def\csname #1\endcsname
                              {\relax\ifmmode #2\else\hbox{$#2$}\fi}}
\defemb{cA}{{\mathcal A}}
\defemb{cB}{{\mathcal B}}
\defemb{cC}{{\mathcal C}}
\defemb{cD}{{\mathcal D}}
\defemb{cE}{{\mathcal E}}
\defemb{cF}{{\mathcal F}}
\defemb{cG}{{\mathcal G}}
\defemb{cH}{{\mathcal H}}
\defemb{cI}{{\mathcal I}}
\defemb{cJ}{{\mathcal J}}
\defemb{cK}{{\mathcal K}}
\defemb{cL}{{\mathcal L}}
\defemb{cM}{{\mathcal M}}
\defemb{cO}{{\mathcal O}}
\defemb{cP}{{\mathcal P}}
\defemb{cS}{{\mathcal S}}
\defemb{cQ}{{\mathcal Q}}
\defemb{cR}{{\mathcal R}}
\defemb{cU}{{\mathcal U}}
\defemb{cV}{{\mathcal V}}
\defemb{cW}{{\mathcal W}}

\newcommand{\var}{{\cV}ar}                
\newcommand{\cs}{~{\leadsto}}                                             
\newcommand{\scs}{~{\stackrel{\mbox{\footnotesize ~SS~}}{\leadsto}}}      
\newcommand{\fcs}{~{\stackrel{\mbox{\footnotesize ~FS~}}{\leadsto}}}      
\newcommand{\ics}{~{\stackrel{\mbox{\footnotesize ~IS~}}{\leadsto}}}      
\newcommand{\W}{~{\Rightarrow}}      
\newcommand{\Wa}{~{\stackrel{\mbox{\footnotesize ~1~}}{\Rightarrow}}}      
\newcommand{\Wc}{~{\stackrel{\mbox{\footnotesize ~3~}}{\Rightarrow}}}      
\newcommand{\Wd}{~{\stackrel{\mbox{\footnotesize ~4~}}{\Rightarrow}}}      

\long\def\comment#1{}

\newcommand{\pr}{{\sc Prolog}}
\newcommand{\ma}{{\sc MALP}}
\newcommand{\fa}{{\sc FASILL}}
\newcommand{\fl}{{\sc FLOPER}}
\newcommand{\BPL}{{\sf Bousi$\sim$Prolog}}
\newcommand{\Llog}{{\sf LIKELOG}}

\newcommand{\SiLog}{{\sf SiLog}}

\title{A Fuzzy Logic Programming Environment\\ 
      for Managing Similarity and Truth Degrees\thanks{This work was supported by the EU (FEDER), and the Spanish MINECO Ministry ({\em Ministerio de Econom\'\i a y Competitividad}) under grant TIN2013-45732-C4-2-P. }}

\author{Pascual Julián-Iranzo
\institute{Department of Technologies and Information Systems\\
 University of Castilla-La Mancha\\
 13071 Ciudad Real (Spain)}
 \email{Pascual.Julian@uclm.es}
\and
Ginés Moreno
\institute{Department of Computing Systems\\
 University of Castilla-La Mancha\\
 02071  Albacete (Spain)}
 \email{Gines.Moreno@uclm.es}
\and Jaime Penabad
\institute{Department of Mathematics\\
 University of Castilla-La Mancha\\
 02071  Albacete (Spain)}
\email{~~~~~~~~~Jaime.Penabad@uclm.es}
~~~~~~~~~~~~~~~~~\and Carlos Vázquez
\institute{Department of Computing Systems\\
 University of Castilla-La Mancha\\
 02071  Albacete (Spain)}
 \email{Carlos.Vazquez@uclm.es}}

\def\titlerunning{A Fuzzy Logic Programming Environment for Managing Similarity and Truth Degrees}
\def\authorrunning{P. Julián-Iranzo, G. Moreno, J. Penabad \& C. Vázquez}

\begin{document}
\maketitle

\begin{abstract}\fa\ (acronym of 
``Fuzzy Aggregators and Similarity Into a Logic Language'') 
is a fuzzy logic programming language with implicit/explicit truth degree annotations, a great variety of connectives and unification by similarity. \fa\ integrates and extends features coming from MALP ({\em Multi-Adjoint Logic Programming}, a fuzzy logic
 language with explicitly annotated rules) 
and Bousi$\sim$Prolog (which uses a weak unification algorithm and is well suited for flexible query answering). Hence, it properly manages similarity and truth degrees in a single framework combining the expressive benefits of both languages.
This paper presents the main features and implementations details of \fa.
Along the paper we describe its syntax and operational semantics and we give clues of the implementation of the lattice module and the similarity module, two of the main building blocks of the new programming environment
which enriches the \fl\ system developed in our research group.
\end{abstract}


\section{Introduction}
\label{sec-syn}

The challenging research area 
of {\em Fuzzy Logic Programming} is 
devoted to introduce {\em fuzzy logic} concepts  into {\em logic programming}
in order to explicitly treat with uncertainty in a natural way. 
It has provided a wide variety of \pr\ dialects along the last three decades. {\em Fuzzy logic languages} can be classified (among other criteria) regarding the emphasis they assign when fuzzifying the original unification/resolution mechanisms of \pr. So, whereas some approaches are able to cope with similarity/proximity  relations at unification time \cite{FGS00,
FA02,Ses02}, other ones extend their operational principles (maintaining syntactic unification) for managing a wide variety of fuzzy connectives and truth degrees on rules/goals beyond the simpler case of {\em true} or {\em false}
\cite{KS92,MOV04,MPH11INS}. 

The first line of integration, where the syntactic unification algorithm is extended with the ability of managing similarity/proximity relations, is of special relevance for this work. Similarity/proximity relations put in relation the elements of a set with a certain approximation degree and serve for weakening the notion of equality and, hence, to deal with vague information. With respect to this line, the related work can be summarized as follows: 

Firstly, the pioneering papers \cite{FFG96,FGS99,FGS00} and \cite{FF02}, where the concept 
of unification by similarity was first developed. Note that, we share their objectives, using similarity relations as a basis, but contrary to our proposal, they use the sophisticated (but cumbersome) notions of {\em clouds}, {\em systems of clouds} and {\em closure operators} in the definition of the unification algorithm, that may endanger the efficiency of the derived operational semantics.

More closely tied to our proposal, is the work presented in  \cite{Ses02} by Maria Sessa. She defines 
an extension of the SLD-resolution principle, incorporating a similarity-based unification procedure which is a reformulation of Martelli and Montanari's unification algorithm \cite{MM82} where symbols match if they are similar (instead of syntactically equal).  
The resulting algorithm uses a generalized notion of most general unifier that provides 
a numeric value, which gives a measure of the approximation degree, and a graded notion of logical 
consequence. Sessa's approach to unification can be considered our starting point. 

From a practical point of view, similarity-based approaches have produced three main 
experimental realizations. The first two system prototypes described in the literature were:
the fuzzy logic language \Llog\ (\emph{LIKEness in LOGic}) \cite{FF99} (an interpreter implemented 
in \pr\ using rather direct techniques and the aforementioned cloud and closure concepts described in \cite{FF02,FFG96,FGS00,FGS99}) and  
\SiLog\ \cite{LSS01} (an interpreter written in Java based on the ideas introduced in \cite{Ses02}). 
Neither \Llog\ nor \SiLog\ are publicly available, what prevent a real evaluation of these systems, 
and they seem immature prototypes. In this same line of work, \BPL\ \cite{JR09IWANN,JRG09ENTCS}, on the other hand, is the first fuzzy logic programming system which is a true \pr\ extension and not a simple interpreter able to execute a weak SLD-resolution procedure.
Also it is the first fuzzy logic programming language that proposed the use of proximity relations as a generalization of similarity relations \cite{JR09PPDP}.
It is worth saying that, in order to deal with proximity relations, \BPL\ has needed to develop new theoretical \cite{JR11IWANN} and conceptual \cite{RJ14JIFS} basis. 

A related programming framework, akin to Fuzzy Logic Programming, is {\em Qualified Logic Programming} 
(QLP) \cite{RR08FLOPS}, which is a derivation of van Emden's {\em Quantitative Logic Programming} 
\cite{Emd86} and {\em Annotated Logic Programming} \cite{KS92}. In QLP a qualification domain $D$ is associated to a program and their rules annotated with qualification values, resulting a parametric framework: QLP(D). In \cite{CRR08} they introduce similarity relations in their QLP(D) framework by adopting a transformational approach. The new Similarity-based QLP(D) scheme, named SQLP(D), transforms a similarity relation into a set of QLP(D) rules able to emulate a unification by similarity process.
In \cite{RR10,CRR14TPLP} they go a step further integrating constraints and proximity relations in their generic scheme, obtaining a really flexible programming framework named SQCLP. 

Ending this section, it is important to say that our research group has been involved  
both on the development of similarity-based logic programming systems and those that extend the resolution principle, as reveals
the design of the Bousi$\sim$Prolog language\footnote{Two different programming environments for Bousi$\sim$Prolog are available at \url{ http://dectau.uclm.es/bousi/}.
}
\cite{JR09PPDP,JR10FUZZIEEE,RJ14JIFS}, where clauses cohabit with similarity/proximity equations,  and the development of the \fl\ system\footnote{The tool is freely accessible from the Web site \url{ http://dectau.uclm.es/floper/}.
}, which manages fuzzy programs composed by rules richer than clauses \cite{MMPV10RULEML,MV14JSEA}.
Our unifying approach is somehow inspired by \cite{CRR14TPLP}, but in our framework we admit a wider set of connectives inside the body of programs rules. In this paper,  we make a first step in our pending task 
of embedding into \fl\ the {\em weak unification} algorithm of Bousi$\sim$Prolog.

The structure of this paper is as follows. Firstly, in Sections \ref{sec-fas} and \ref{sec-mal} we formally define and illustrate both the syntax and operational semantics, respectively, of the \fa\ language. Next, Section \ref{sec-imp} is concerned with implementation and practical issues. Finally, in Section \ref{sec-con} we conclude by proposing too further research.

\section{The \fa\ language}\label{sec-fas}

\fa\ is a first order language built upon a signature $\Sigma$, that contains the elements of a countably infinite set of variables $\cV$, function symbols and predicate symbols with an associated arity --usually expressed as pairs $f/n$ or $p/n$ where $n$ represents its arity--, 
the implication symbol ($\leftarrow$) and a wide set of others connectives. The language combines the elements of $\Sigma$ as terms, atoms, rules and formulas.  A \emph{constant} $c$ is a function symbol with arity zero. A {\it term} is a variable, a constant or a function symbol $f/n$ applied to $n$ terms $t_1,\ldots,t_n$, and 
is
denoted as $f(t_1,\ldots,t_n)$. We allow values of a lattice $L$ as part of the signature $\Sigma$. Therefore, a well-formed formula can be either:
\begin{itemize}
\item $r$, if $r\in L$
\item $p(t_1,\ldots,t_n)$, if $t_1,\ldots,t_n$ are terms and $p/n$ is an n-ary predicate. This formula is called \emph{atom}. Particularly, atoms containing no variables are called \emph{ground atoms}, and atoms built from nullary predicates are called \emph{propositional variables}
\item $\varsigma(\cF_1,\ldots,\cF_n)$, if $\cF_1,\ldots,\cF_n$ are well-formed formulas and $\varsigma$ is an n-ary connective with truth function $\dot{\varsigma}:L^n\rightarrow L$
\end{itemize}

\begin{definition}[Complete lattice]
A complete lattice is a partially ordered set $(L,\leq)$ such that every subset $S$ of $L$ has infimum and supremum elements. Then, it is a bounded lattice, i.e., it has bottom and top elements, denoted by $\bot$ and $\top$, respectively. $L$ is said to be the {\it carrier set} of the lattice, and $\leq$ its ordering relation.
\end{definition}

\noindent The language is equipped with a set of {\it connectives}\footnote{Here, the connectives are binary operations but we usually generalize them with an arbitrary number of arguments.} interpreted on the lattice, including 
\begin{itemize}
\item aggregators denoted by $@$, whose truth functions $\dot{@}$  fulfill
the boundary condition: \hspace{-0.1cm}$\dot{@}(\top,\hspace{-0.05cm}\top)$  $= \top$, $\dot{@}(\bot,\bot)=\bot$, and monotonicity: 
$(x_1,y_1)\leq (x_2,y_2)\Rightarrow\dot{@}(x_1,y_1)\leq\dot{@}(x_2,y_2)$. 
\item t-norms and t-conorms \cite{NW06} (also named conjunctions and disjunctions, that we denote by $\&$ and $|$, respectively) whose truth functions fulfill the following properties:

\begin{tabular}{lll}
$\cdot$ Commutative: &\hspace{-3cm}$\dot{\&}(x,y) = \dot{\&}(y,x)$                         &\hspace{-3.3cm} $\dot{|}(x,y)=\dot{|}(y,x)$\\
$\cdot$ Associative: &\hspace{-3cm}$\dot{\&}(x,\dot{\&}(y,z)) = \dot{\&}(\dot{\&}(x,y),z)$ &\hspace{-3.3cm} $\dot{|}(x,\dot{|}(y,z)) = \dot{|}(\dot{|}(x,y),z)$\\
$\cdot$ Identity element: &\hspace{-3cm}$\dot{\&}(x,\top) = x$                             &\hspace{-3.3cm} $\dot{|}(x,\bot) = x$\\
$\cdot$ Monotonicity in each argument: $z\leq t\Rightarrow$                  &
                                                         \hspace{-0.35cm}$\left\{\begin{array}{ll}
                                                         \dot{\&}(z,y) \leq \dot{\&}(t,y)&\dot{\&}(x,z)\leq\dot{\&}(x,t)\\
																					               \dot{|}(z,y) \leq  \dot{|}(t,y) & \dot{|}(x,z) \leq \dot{|}(x,t)\\
																					               \end{array}
																					               \right.
																					               $
\end{tabular}
\end{itemize}

\begin{example}\label{exa-lat}
\noindent 
In this paper we use 
the lattice $([0,1],\leq)$, where $\leq$ is the usual ordering relation on real numbers, 
and three sets of connectives corresponding to the fuzzy logics of G\"{o}del, \L{}ukasiewicz and Product, defined in Figure \ref{F0}, where labels $\tt L$, $\tt G$ and $\tt P$ mean respectively {\em
{{\L{}}}ukasiewicz logic}, {\em G\"{o}del logic} and {\em product
logic} (with different capabilities for modeling {\em pessimistic},
{\em optimistic} and {\em realistic scenarios}).

It is possible to include also other connectives. For instance, the arithmetical average, defined by connective $@_{aver}$ (with truth function $\dot{@}_{aver}(x,y) \triangleq \frac{x+y}{2}$), that is a stated, easy to understand connective that does not belong to a known logic. Connectives with arities different from 2 can also be used, like the $@_{very}$ aggregation, defined by $\dot{@}_{very}(x) \triangleq x^2$, that is a unary connective.

\end{example}

\begin{figure}[t] $ 
~~~~~~~~~~~~~~~~~~~~~~~~~~\begin{array}{lllllllllrl} 
 \dot{\&}_{\tt P} (x,y) \triangleq x *  y &
  \dot{|}_{\tt P}(x,y) \triangleq x+y-xy & ~
  \hbox{\emph{Product}}\\
  \dot{\&}_{\tt G} (x,y) \triangleq \min(x,y) &
  \dot{|}_{\tt G}(x,y) \triangleq  max(x,y)   & ~
      \hbox{\emph{G\"odel }}\\
  \dot{\&}_{\tt L} (x,y)  \triangleq \max(0,x+y-1)  &
  \dot{|}_{\tt L}(x,y)  \triangleq  \min(x+y,1) & ~ 
    \hbox{\emph{\L ukasiewicz}}
  \end{array}
$ \caption{Conjunctions and disjunctions in $[0,1]$ for {\em Product}, {\em {{\L{}}}ukasiewicz}, and 
{\em G\"{o}del} fuzzy  logics}\label{F0}
\end{figure}

\begin{definition}[Similarity relation]
Given a domain $\cU$ and a lattice $L$ with a fixed t-norm $\wedge$, a similarity relation $\cR$ is a fuzzy binary relation on $\cU$,
that is a fuzzy subset on $\cU\times\cU$ (namely, a mapping $\cR : \cU \times\cU \rightarrow L$), such that fulfils the following properties\footnote{For convenience, $\cR(x,y)$, also denoted $x\cR y$, refers to both the syntactic expression (that symbolizes that the elements $x,y\in\cU$ are related by $\cR$) and the truth degree $\mu_\cR(x,y)$, i.e., the affinity degree of the pair $(x,y)\in\cU\times\cU$ with the verbal predicate $\cR$.}:
\begin{itemize}
\item Reflexive: $\cR(x,x) = \top, \forall x\in\cU$
\item Symmetric: $\cR(x,y) = \cR(y,x), \forall x,y\in\cU$
\item Transitive: $\cR(x, z) \geq \cR(x, y) \wedge \cR(y, z), \forall x,y,z\in\cU$
\end{itemize}

\end{definition}
\noindent Certainly, we are interested in fuzzy binary relations on a syntactic domain. We primarily define
similarities on the symbols of a signature, $\Sigma$, of a first order language. This makes possible to
treat as indistinguishable two syntactic symbols which are related by a similarity relation $\cR$.
Moreover, a similarity relation $\cR$ on the alphabet of a first order language can be extended to terms
by structural induction in the usual way \cite{Ses02}.
That is, the extension, $\hat\cR$, of a similarity relation $\cR$ is defined as:
\begin{enumerate}
\item let $x$ be a variable, $\hat\cR(x,x)=\cR(x,x)=1$,

\item let $f$ and $g$ be two $n$-ary function symbols and let
$t_1$, \ldots, $t_n$, $s_1$, \ldots, $s_n$ be terms, \\[1ex]
$\hat\cR(f(t_1, \ldots, t_n),g(s_1, \ldots, s_n)) =
\cR(f,g) \wedge (\bigwedge_{i=1}^n \hat\cR(t_i,s_i))$

\item otherwise, the approximation degree of two terms is zero.
\end{enumerate}
Analogously for atomic formulas. Note that, in the sequel, we shall not make a notational distintion
between the relation $\cR$ and its extension $\hat\cR$.

\begin{example}\label{exa-sim}
A similarity relation $\cR$ on the elements of $\cU=\{vanguardist, elegant, metro, taxi, bus\}$ is defined by the following matrix: 

\begin{tabular}{lll}
\begin{tabular}{l||c|c|c|c|c|}
$\cR$        & vanguardist & elegant & metro & taxi & bus \\
\hline
\hline
vanguardist  &       1     &   0.6   &   0   &   0  &  0  \\
\hline
elegant      &      0.6    &    1    &   0   &   0  &  0  \\
\hline 
metro        &       0     &    0    &   1   &  0.4 &  0.5\\
\hline 
taxi         &       0     &    0    &   0.4 &   1  &  0.4\\
\hline
bus          &       0     &    0    &   0.5 &  0.4 &  1\\
\hline
\end{tabular}
& ~ &
\parbox{4.85cm}{
It is easy to check that $\cR$ fulfills the reflexive, symmetric and transitive properties. 
Particularly, using the \emph{G\"odel} conjunction as the t-norm $\wedge$, we have that: 
$\cR(taxi, metro) \geq \cR(metro, bus) \wedge \cR(bus, taxi) = 0.5 \wedge 0.4$.
}\\[1.5ex]
\end{tabular}

\noindent Furthermore, the extension $\hat\cR$ of $\cR$ determines that the terms 
$elegant(taxi)$ and $vanguardist(metro)$ are similar, since: 
$\hat{\cR}(elegant(taxi),vanguardist(metro)) = 
   \cR(elegant, vanguardist) \wedge \hat{\cR}(taxi, metro) = 
   0.6 \wedge {\cR}(taxi, metro) = 
   0.6 \wedge 0.4 = 0.4$.
\end{example}

\begin{definition}[Rule]
A {\it rule} has the form $A\leftarrow\cB$, where $A$ is an atomic formula called {\it head} and $\cB$, called {\it body}, is a well-formed formula (ultimately built from atomic formulas $B_1,\ldots,B_n$, truth values of $L$ and connectives)
\footnote{In order to subsume the syntactic conventions of \ma, in our 
programs we also admit {\it weighted rules}
with shape ``$A\leftarrow_i\cB ~with~ v$'', which are internally treated as ``$A\leftarrow (v\&_i\cB)$'' (this transformation preserves the meaning of rules
as proved in \cite{MPV13CMMSE}).}.
In particular, when the body of a rule is $r\in L$ (an element of lattice $L$), this rule is called {\it fact} and can be written as $A\leftarrow r$ (or simply $A$ if $r=\top$).
\end{definition}

\begin{definition}[Program]
A program $\cP$ is a tuple $\tuple{\Pi,\cR,L}$ where $\Pi$ is a set of rules, $\cR$ is a similarity relation between the elements of $~\Sigma$, and $L$ is a complete lattice.
\end{definition}

\begin{example}\label{exa-pro}
The set of rules $\Pi$ given below, the similarity relation $\cR$ of Example \ref{exa-sim} and lattice $L = ([0,1],\leq)$ of Example \ref{exa-lat}, form a program
$\cP = \tuple{\Pi,\cR,L}$.

$$
\Pi\left\{
\begin{array}{lll}
R_1:&vanguardist(hydropolis) & \leftarrow 0.9\\
R_2:&elegant(ritz) & \leftarrow 0.8\\
R_3:&close(hydropolis,taxi) & \leftarrow 0.7\\
R_4:&good\_hotel(x) & \leftarrow @_{aver}(elegant(x), @_{very}(close(x, metro)))
\end{array}\right.
$$
\end{example}

\section{Operational Semantics of \fa}\label{sec-mal}

\noindent Rules in a \fa\ program have the same role than clauses in \pr\ (or \ma\ \cite{MOV04,JMP09IWANN,MPV14JRS}) programs, that is, stating that a certain predicate relates some terms (the {\it head}) if some conditions (the {\it body}) hold.

As a logic language, \fa\ inherits the concepts of substitution, unifier and most general unifier ($mgu$). Some of them are extended to cope with similarities. 
Concretely, following the line of Bousi$\sim$Prolog \cite{JR09PPDP}, the most general unifier is replaced by the concept of {\it weak most general unifier} (w.m.g.u.) and a weak unification algorithm is introduced to compute it.
Roughly speaking, the {\em weak unification algorithm} states that two {\em expressions} (i.e, terms or
atomic formulas) $f(t_{1},\ldots,t_{n})$ and $g(s_{1},\ldots,s_{n})$ weakly unify if the root symbols
$f$ and $g$ are close with a certain degree 
(i.e. $\cR(f,g)=r>\bot$) and each of their arguments $t_i$ and $s_i$ weakly unify.
Therefore, there is a weak unifier for two expressions even if the symbols at their roots are not
syntactically equals ($f\not\equiv g$).

More technically,
the weak unification algorithm we are using is a reformulation/extension
of the one which appears in \cite{Ses02} for arbitrary complete lattices. We formalize it
as a transition system supported by a similarity-based unification relation ``$\Rightarrow$''.
The unification of the expressions $\cE_1$ and $\cE_2$ is obtained by a state transformation
sequence starting from an initial state
$\tuple{G\equiv \{\cE_1\approx \cE_2\}, id, \alpha_0}$,
where $id$ is the identity substitution and $\alpha_0=\top$ is the supreme of $(L,\leq)$:
$
\tuple{G, id, \alpha_0} \Rightarrow \tuple{G_1, \theta_1, \alpha_1} \Rightarrow
\cdots \Rightarrow
\tuple{G_n, \theta_n, \alpha_n}.~
$
When the final state $\tuple{G_n, \theta_n, \alpha_n}$, with
$G_n= \emptyset$, is reached (i.e., the equations in the initial
state have been solved),
the expressions $\cE_1$ and $\cE_2$ are unifiable by similarity with
w.m.g.u. $\theta_n$ and {\em unification degree} $\alpha_n$.
Therefore, the final state $\tuple{\emptyset, \theta_n, \alpha_n}$
signals out the unification success. On the other hand, when
expressions $\cE_1$ and $\cE_2$ are not unifiable, the state transformation
sequence ends with failure (i.e., $G_n = Fail$).

The \emph{similarity-based unification relation},  ``$\W$'', is defined as the
smallest relation derived by the following set of transition rules (where $\var(t)$ denotes the set of variables of a given term $t$)
\begin{center}

$\inferrule*[Right=1]
    {\tuple{\{f(t_1,\ldots,t_n)\approx g(s_1,\ldots,s_n)\}\cup E,\theta,r_1} \\ \cR(f,g)=r_2 > \bot}
    {\tuple{\{t_1\approx s_1,\ldots,t_n\approx s_n\}\cup E,\theta,r_1\wedge r_2}}$\\[0.3cm]

\begin{tabular}{ccccc}
$\inferrule*[Right=2]
    {\tuple{\{X\approx X\}\cup E,\theta,r_1}}
    {\tuple{E,\theta,r_1}}$& & & &

$\inferrule*[Right=3]
    {\tuple{\{X\approx t\}\cup E,\theta,r_1} \\ X\notin \var(t)}
    {\tuple{(E)\{X/t\},\theta\{X/t\},r_1}}$
\\[0.3cm]
$\inferrule*[Right=4]
    {\tuple{\{t\approx X\}\cup E,\theta,r_1}}
    {\tuple{\{X\approx t\}\cup E,\theta,r_1}}$& & & &

$\inferrule*[Right=5]
    {\tuple{\{X\approx t\}\cup E,\theta,r_1} \\ X\in \var(t)}
    {\tuple{Fail,\theta,r_1}}$\\[0.3cm]
\end{tabular}

$\inferrule*[Right=6]
    {\tuple{\{f(t_1,\ldots,t_n)\approx g(s_1,\ldots,s_n)\}\cup E,\theta,r_1} \\ \cR(f,g)= \bot}
    {\tuple{Fail,\theta,r_1}}$

\end{center}

\noindent Rule $1$ decomposes two expressions and annotates the relation between 
the
function (or predicate) symbols 
at their root.
The second rule eliminates spurious information and the fourth rule interchanges
the position of the symbols to be handled by other rules. The third and fifth rules perform an occur check of variable $X$ in a term $t$. In case of success, it generates a substitution $\{X/t\}$; otherwise the algorithm ends with failure. It can also end with failure if the relation between function (or predicate) symbols in $\cR$ is $\bot$, as stated 
by
Rule $6$.

Usually, given two expressions $\cE_1$ and $\cE_2$, if there is a successful transition sequence,
$\tuple{\{\cE_1\approx \cE_2\},id,\top}\Rightarrow^{\star} \tuple{\emptyset,\theta,r}$, then we write that
$wmgu(\cE_1,\cE_2) = \tuple{\theta,r}$, being $\theta$ the {\it weak most general unifier} of $\cE_1$
and $\cE_2$, and $r$ is their {\it unification degree}.

Finally note that, in general, a w.m.g.u. of two expressions $\cE_1$ and $\cE_2$ is not unique \cite{Ses02}.
Certainly, the weak unification algorithm only computes a representative of a w.m.g.u. class, in the 
sense that, if $\theta = \{x_1/t_1, \ldots, x_n/t_n\}$ is a w.m.g.u., with 
degree $\beta$, 
then, by definition, any substitution 
$\theta' = \{x_1/s_1, \ldots, x_n/s_n\}$, satisfying $\cR(s_i,t_i) > \bot$, 
for any $1\leq i \leq n$, is also a w.m.g.u. with approximation degree 
$\beta'= \beta\wedge (\bigwedge_{1}^{n} \cR(s_i,t_i))$, where ``$\wedge$'' is a selected t-norm.
However, observe that, the w.m.g.u. representative computed by the weak unification algorithm is
one with an approximation degree equal or greater than any other w.m.g.u. As in the case of the classical syntactic unification algorithm, our algorithm always terminates returning a success or a failure.

Next, we illustrate the weak unification process in the following example.

\begin{example}\label{exa-wmg}
Consider the lattice $L=([0,1],\leq)$ of Example \ref{exa-lat} and the relation $\cR$ of Example \ref{exa-sim}. Given terms $elegant(taxi)$ and $vanguardist(metro)$, it is possible the following weak unification process:
\[
\begin{array}{l}
\tuple{\{elegant(taxi) \approx vanguardist(metro)\}, id, 1} \Wa 
      \tuple{\{taxi\approx metro\}, id, 0.6}  \Wa \\
\tuple{\{\}, id, 0.6 \wedge 0.4} =
      \tuple{\{\}, id, 0.4}
\end{array}
\]

\noindent Also it is possible the unification of the terms $elegant(taxi)$ and $vanguardist(X)$, since:
\[
\begin{array}{l}
\tuple{\{elegan(taxi) \approx vanguardist(X)\}, id, 1} \Wa
      \tuple{\{taxi\approx X\}, id, 0.6} \Wd \\
\tuple{\{X\approx taxi\}, id, 0.6} \Wc
      \tuple{\{\}, \{X/taxi\}, 0.6} 
\end{array}
\]
and the substitution $\{X/taxi\}$ is their w.m.g.u. with unification degree $0.6$.
\end{example}

In order to describe the procedural semantics of the
 \fa\  language, in the following we denote by $\cC[A]$ a formula
where $A$ is a sub-expression (usually an atom) which occurs in
the --possibly empty-- context $\cC[]$ whereas $\cC[A/A']$ means
the replacement of $A$ by $A'$ in the
context $\cC[]$. Moreover,
$\var(s)$ denotes the set of distinct variables occurring in the
syntactic object $s$ 
and $\theta [\var(s)]$ refers to  the
substitution obtained from $\theta$ by restricting its domain to
$\var(s)$. 
In the next definition, we always consider
that $A$ is the selected atom in 
a
goal $\cQ$ and $L$ is the complete lattice associated to $\Pi$.
\begin{definition} [Computational Step] \label{ad}
Let $\cQ$ be a goal and let $\sigma$ be a substitution. The pair
$\tuple{\cQ; \sigma}$ is a {\em state}. Given a program $\tuple{\Pi,\cR, L}$ and a t-norm $\wedge$ in $L$, a \emph{computation} is formalized as a state transition
system, whose transition relation $\cs{}$ is the smallest relation
satisfying 
these rules:\\
\noindent 1) \emph{Successful step} (denoted as $\scs$):
\begin{center}
$\inferrule*[Right=SS] {\tuple{\cQ[A],\sigma} \\ A'\leftarrow\cB\in\Pi \\ wmgu(A,A')=\tuple{\theta,r}}
                        {\tuple{\cQ[A/\cB\wedge r]\theta, \sigma\theta}}$\\[0.3cm]
\end{center}
\noindent 2) \emph{Failure step} (denoted as $\fcs$):
\begin{center}
$\inferrule*[Right=FS] {\tuple{\cQ[A],\sigma} \\ \nexists A'\leftarrow\cB\in\Pi : wmgu(A,A')=\tuple{\theta, r}, r>\bot}
                        {\tuple{\cQ[A/\bot], \sigma}}$
\end{center}
\noindent 3) \emph{Interpretive step} (denoted as $\ics$):
\begin{center}
$\inferrule*[Right=IS]  {\tuple{\cQ[@(r_1,\ldots,r_n)];\sigma} \\ \dot{@}\!(r_1,\ldots,r_n)=r_{n+1}}
                        {\tuple{\cQ[@(r_1,\dots,r_n)/r_{n+1}]; \!\sigma}}$\\[0.3cm]
\end{center}
\end{definition}

\noindent A \emph{derivation} is a sequence of arbitrary lenght $\tuple{\cQ;
id}\cs{^*} \tuple{\cQ';\sigma}$. As usual, rules are 
renamed apart. When
$\cQ' = r\in L$, 
the state $\tuple{r;\sigma}$ is called a {\em fuzzy computed
answer} (f.c.a.) for that derivation.

\begin{figure*}[t]
	\center
   \label{fig-program-goal}
   \includegraphics[width=0.9\textwidth]{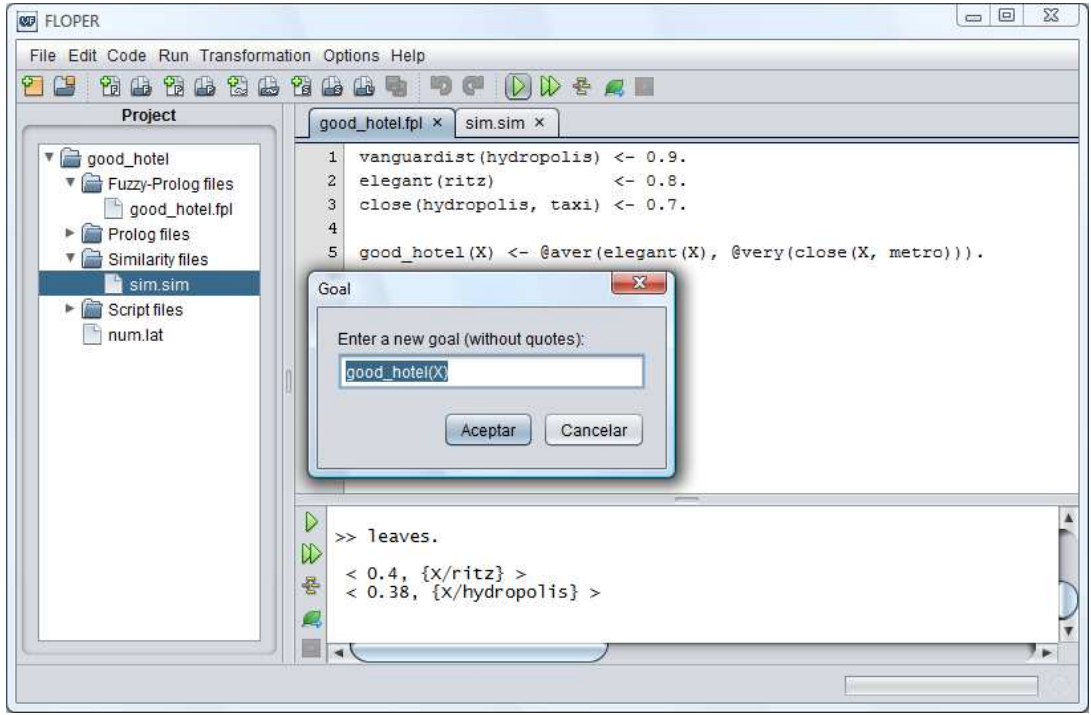} 
   \caption{Screen-shot of a work session with
	\fl\ managing a \fa\ program}
\end{figure*}

\begin{example}\label{exa-opr}
Let $\cP = \tuple{\Pi,\cR,L}$ be the program from Example \ref{exa-pro}, and $\cQ=good\_hotel(X)$ be a goal. 
It is possible to perform these two derivations for $\cP$ and $\cQ$:

{\small
$
\begin{array}{lll}
D_1: & \tuple{good\_hotel(X), id}					& \scs^{R4} \\
	&		 \tuple{@_{aver}(elegant(X), @_{very}(close(X,metro))), \{X_1/X\}}	& \scs^{R2} \\
	&		 \tuple{@_{aver}(0.8, @_{very}(close(ritz,metro))), \{X_1/ritz,X/ritz\}}	& \fcs \\
	&		 \tuple{@_{aver}(0.8, @_{very}(0)), \{X_1/ritz,X/ritz\}}	& \ics \\
	&		 \tuple{@_{aver}(0.8, 0), \{X_1/ritz,X/ritz\}}	& \ics \\[0.3ex]
	&		 \tuple{0.4, \{X_1/ritz,X/ritz\}} & \\
\end{array}
$
}

{\small
$
\begin{array}{lll}
D_2: & \tuple{good\_hotel(X), id}					& \scs^{R4} \\
     & \tuple{@_{aver}(elegant(X), @_{very}(close(X,metro))), \{X_1/X\}}	& \scs^{R1} \\
	&		 \tuple{@_{aver}(\&_{godel}(0.9,0.6), @_{very}(close(hydropolis,metro))), \{X_1/hydropolis, X/hydropolis\}}	& \scs^{R3} \\
	&		 \tuple{@_{aver}(\&_{godel}(0.9,0.6), @_{very}(\&_{godel}(0.7,0.4))), \{X_1/hydropolis, X/hydropolis\}}	& \ics \\
		&	 \tuple{@_{aver}(0.6, @_{very}(0.4)), \{X_1/hydropolis, X/hydropolis\}}	& \ics \\
		&	 \tuple{@_{aver}(0.6, 0.16), \{X_1/hydropolis, X/hydropolis\}}	& \ics \\[0.3ex]
		&	 \tuple{0.38, \{X_1/hydropolis, X/hydropolis\}}	&  \\
\end{array}
$\\
}

\noindent
with fuzzy computed answers 
$\tuple{0,4, \{X/ritz\}}$ and 
$\tuple{0.38, \{X/hydropolis\}}$, respectivelly.
\end{example}

\begin{figure*}[t] \vspace{-0.2cm}
   \label{fig-tree}
   \center
	\includegraphics[width=0.75\textwidth]{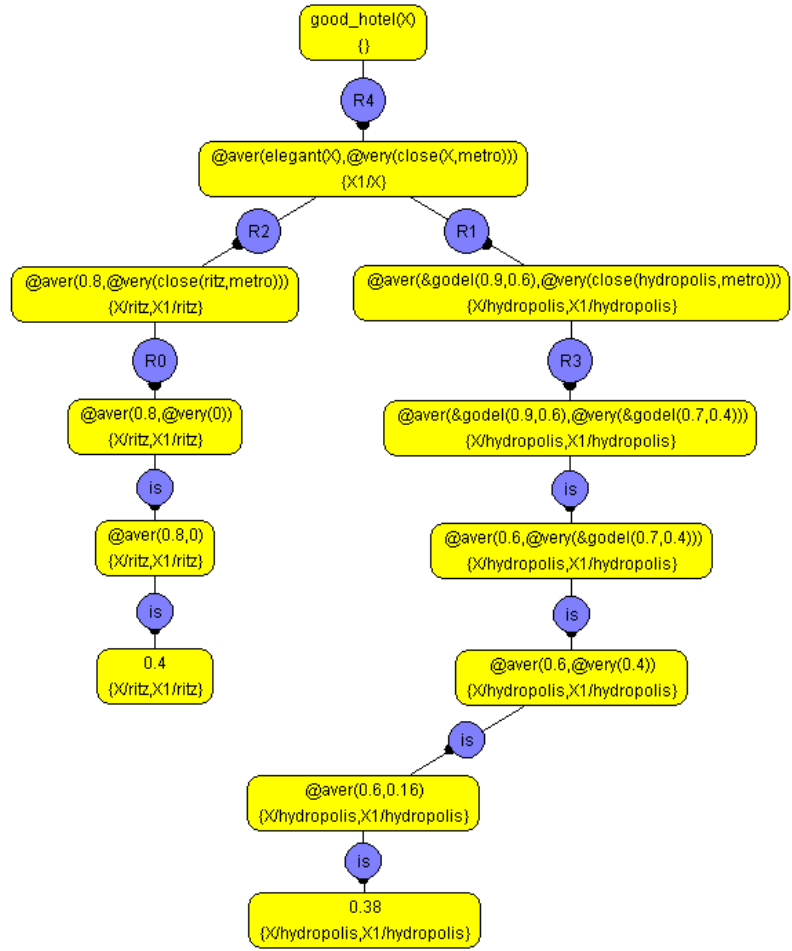} 
   \caption{An execution tree as shown by the \fl\ system}
\end{figure*}

\section{Implementation of \fa\ in \fl}\label{sec-imp}

During the last years we have developed the \fl\ tool, initially intended for manipulating \ma\ programs\footnote{ 
The \ma\ language is nowadays fully subsumed by the new \fa\ language 
just introduced in this paper, since, given a \fa\ program $\cP=\tuple{\Pi,\cR,L}$, if $\cR$ is 
the identity relation (that is, the one 
where each element of a signature $\Sigma$ is only similar to itself, with the maximum similarity degree) and $L$ is a complete lattice also containing {\em adjoint pairs} \cite{MOV04}, then $\cP$ is a \ma\ program too.
}. In its current development state, \fl\ has been equipped with new features in order to cope with more expressive languages and, in particular, with \fa. 
The new version of \fl\ is freely accessible in the URL \url{ http://dectau.uclm.es/floper/?q=sim} where it is possible to test/download the new prototype incorporating the management of similarity relations. In this section we briefly describe the main features of this tool before presenting the novelties introduced in this work. 

\fl\ has been implemented in Sicstus Prolog v.3.12.5 (rounding about 1000 
lines of code, where our last update supposes approximately a 30\% of the final code) and it has been recently equipped with a graphical interface written in Java (circa 2000 lines of code). More detailed, the \fl\ system consists in a JAR (Java archive) file that runs the graphical interface. This JAR file calls a \pr\ file containing the two main independent blocks: 1) the Parsing block parses \fa\ files into two kinds of \pr\ code (a high level platform-independent \pr\ program and a set of facts to be used by \fl), and 2) the Procedural block performs the evaluation of a goal against the program, implementing the procedural semantics previously described. This code is completed with a configuration file indicating the location of the \pr\ interpreter as well as some other data.

\fl\ provides a traditional command interpreter. When the command interpreter is executed, it
offers a menu with a set of commands grouped in four submenus:

\begin{itemize}
\item ``Program Menu'': includes options for {\it parsing} a \fa\ program from a file with extension ``{\tt .fpl}'', {\it saving} the generated \pr\ code to a ``{\tt .pl}'' file, {\it loading/parsing} a pure \pr\ program, {\it listing} the rules of the parsed program and {\it cleaning} the database.
\item ``Lattice Menu'': allows the user to change and show the lattice (implemented in \pr) associated to a fuzzy program through options {\it lat} and  {\it show}, respectively.
\item ``Similarity Menu'': option {\it sim} allows the user to load a similarity file (with extension ``{\tt .sim}'', and whose syntax is detailed further in 
the Similarity Module subsection
) and {\it tnorm} sets the conjunction to be used in the transitive closure of the relation.
\item ``Goal Menu'': by choosing option {\it intro} the user introduces the goal to be evaluated.
Option {\it tree} draws the execution tree for that goal whereas {\it leaves} only shows the 
fuzzy computed answer
contained on it, and {\it depth} is used for fixing its maximum depth. 

\end{itemize}

\noindent The syntax of \fa\ presented in Section \ref{sec-syn} is easily translated to be written by a computer. As usual in logic languages, variables are written as 
identifiers
beginning by an upper case character or an underscore ``{\tt \_}'', while function and predicate symbols are expressed with 
identifiers
beginning by a lower case character, and numbers are 
literals.
Terms and atoms have the usual syntax (the function or predicate symbol, if no nullary, is followed by its arguments between parentheses and separated by a colon). Connectives are labeled with their name immediately after. The implication symbol is written as ``{\tt <-}'', and each rule ends with a dot. Additionally it is possible to include pure \pr\ expressions inside the body of a rule by encapsuling them between curly brackets ``{\tt \{\}}'', and \pr\ clauses between the dollar symbol ``{\tt \$}'', together with \fa\ rules.

The graphical interface (written in Java)
supports a friendly interaction with the user,
as seen in Figure \ref{fig-program-goal}. The graphical interface shows three areas. The leftmost one draws the project tree (grouping each category of file into its own directory). In the right part, the upper area displays the selected file of the tree and the lower one shows the code and the solutions of executing a goal. This interface groups files into projects
which 
include a set of \emph{fuzzy} files ({\tt .fpl}), \pr\ files ({\tt .pl}), \emph{ similarity} files ({\tt .sim}), \emph{script} files
 -containing a  list of commands to be executed consecutively-
({\tt .vfs})
and just one lattice file ({\tt .lat}). When executing a goal, the tool considers the whole program merged from the set of files, thus obtaining only one fuzzy program, one similarity relation, one lattice and one \pr\ file.

\paragraph{The lattice module.}
Lattices are described in a {\tt .lat} file using a language that is a subset of \pr\ where the definition of some predicates are mandatory, and the definition of aggregations follows a certain syntax. The mandatory predicates are {\tt member/1}, that identifies the elements of the lattice, {\tt bot/1} and {\tt top/1}, that stand for the infimum and supremum elements of the lattice, and {\tt leq/2}, that implements the ordering relation. Predicate {\tt members/1}, that returns in a list all the elements of the lattice, is only required if it is finite. Connectives are defined as predicates whose meaning is given by a number of clauses. The name of a predicate has the form ${\tt and\_}label$, ${\tt or\_}label$ or ${\tt agr\_}label$ depending on whether it implements a conjunction, a disjunction or an aggregator, where $label$ is an identifier of that particular connective (this way one can define several conjunctions, disjunctions and other kind of aggregators instead of only one). The arity of the predicate is $n+1$, where $n$ is the arity of the connective that it implements, so its last parameter 
is a variable to be unified with the truth value resulting of its evaluation.
\[\left.\begin{array}{l}
?-\ agr\_label(r_1,\ldots,r_n,R). \\
R=r. \\
\end{array}
\right\} {\tt if\ } @_{label}(r_1,\ldots,r_n) = r
\]

\begin{example}
\noindent For instance, the following
clauses show the \pr\ program modeling the lattice of the real interval
$[0,1]$ with the usual ordering relation and connectives
(conjunction and disjunction of the {\em Product logic}, as well
as the average aggregator):
{\begin{verbatim}
 member(X):- number(X), 0=<X, X=<1.                      leq(X,Y):- X=<Y.                                    
 and_prod(X,Y,Z) :- Z is X*Y.                            bot(0).
 or_prod(X,Y,Z)  :- U1 is X*Y, U2 is X+Y, Z is U2-U1.    top(1).
 agr_aver(X,Y,Z) :- U1 is X+Y, Z is U1/2.
\end{verbatim}
}
\end{example}

\paragraph{The similarity module.}
We describe now the main novelty introduced in the tool, that is the ability to take into account a similarity relation. The similarity relation $\cR$ is loaded from a file with extension {\tt .sim} through option {\it sim}. The relation is represented following a concrete syntax:\\

\noindent
\begin{tabular}{ll}
$\tuple{Relation}$ &::= $\tuple{Sim}$ $\tuple{Relation}$ $|$ $\tuple{Sim}$\\
$\tuple{Sim}$ &     ::= $\tuple{Id_f}$[`/' $\tuple{Int_n}$] `$\sim$' $\tuple{Id_g}$[`/' $\tuple{Int_n}$] `=' $\tuple{r}$ `.' $~~|~~$ `$\sim$' `tnorm' `=' $\tuple{tnorm}$ 
\end{tabular}\\

\noindent The {\it Sim} option parses expressions like ``$f \sim g = r$'', where $f$ and $g$ are propositional variables or constants and $r$ is an element of $L$. It also copes with expressions including arities, like ``$f/n \sim g/n = r$'' (then, $f$ and $g$ are function or predicate symbols). In this case, both arities have to be the same. It is also possible to explicit, through a line like ``$\sim tnorm = \tuple{label}$'' the conjunction to be used further in the 
construction of the
transitive closure of the relation. Internally \fl\ stores each relation as a fact $r$ in an ad hoc module $sim$ as $r(f/n,g/n,r)$, where $n=0$ if it has not been specified (that is, the symbol is considered as a constant). The {\tt .sim} file contains only a small set of similarity equations that \fl\ completes by performing the reflexive, symmetric and transitive closure. The first one simply consists of the assertion of the fact $r(A,A,\top)$. 
The symmetric closure produces, for each $r(a,b,r)$, the assertion of its symmetric entry $r(b,a,r)$ if there is not already some $r(b,a,r')$ where $r\leq r'$ (in this case $r(a,b,r)$ will be rewritten as $r(a,b,r')$ when considering $r(b,a,r)$). The transitive closure is computed by the next algorithm\footnote{ 
It is important to note that this algorithm must be executed right after 
performing  the symmetric, reflexive closure.}, where $\wedge$ stands for the conjunction specified by the directive ``{\it tnorm}'', and ``{\it assert}'' and ``{\it retract}'' are self-explainable and defined as in \pr: \\[0.2cm]
\indent\indent\indent\indent\indent\indent\begin{tabular}{l}
\small
\noindent {\bf Transitive Closure }\\
{\bf forall} r(A,B,$r_1$) {\bf in} {\tt sim}\\
\indent{\bf forall} r(B,C,$r_2$) {\bf in} {\tt sim}\\
\indent\indent$r=r_1\wedge r_2$\\
\indent\indent{\bf if} r(A,C,$r'$) {\bf in} {\tt sim}\ {\bf and} $r'<r$\\
\indent\indent\indent {\bf retract} r(A,C,$r'$) {\bf from} {\tt sim}\\
\indent\indent\indent {\bf retract} r(C,A,$r'$) {\bf from} {\tt sim}\\
\indent\indent{\bf end if}\\
\indent\indent{\bf if} r(A,C,$r'$) {\bf not in} {\tt sim}\\
\indent\indent\indent {\bf assert} r(A,C,$r$) {\bf in} {\tt sim}\\
\indent\indent\indent {\bf assert} r(C,A,$r$) {\bf in} {\tt sim}\\
\indent\indent{\bf end if}\\
\indent{\bf end forall}\\
{\bf end forall}
\end{tabular}\\[0.1cm]

\noindent It is important to note that, it is not relevant if the user provides (apparently) inconsistent 
similarity equations, since \fl\ automatically changes the user values by the appropriate 
approximation degrees in order to preserve the properties of a similarity. For instance, if a 
user provides a set of equations such as, 
$a\sim b = 0.8$, $b \sim c = 0.6$ and $a \sim c = 0.3$,  
after the application of our algorithm for the construction of a similarity, results in the 
set of equations 
$a\sim b = 0.8$, $b \sim c = 0.6$ and $a \sim c = 0.6$,
which positively preserves the transitive property\footnote{
For simplicity, we have omitted the equations obtained during the construction of the reflexive, symmetric  closure.
}.

\begin{example}
Let $L$ be the lattice $([0,1],\leq)$.
To illustrate the enhanced expressiveness of 
\fa,
consider the program $\tuple{\Pi,\cR,L}$ 
that models the concept of \emph{good hotel}, that is, an elegant hotel that is very close to a metro entrance, as seen in Figure \ref{fig-program-goal}.
Here, we use an {\it average aggregator} defined as $ \dot{@}_{avg}(x,~y) \triangleq (x+y)/2$, whereas
\emph{very} is a linguistic modifier implemented as well as an aggregator (with arity 1)
with truth function $\dot{@}_{very} ~x  \triangleq x^2$.
The similarity relation $\cR$ states that \emph{elegant} is similar to \emph{vanguardist}, and \emph{metro} to \emph{bus} and (by transitivity) to \emph{taxi}:
\begin{verbatim}
      ~tnorm  = godel                             metro ~ bus  = 0.5.
      elegant/1 ~ vanguardist/1  = 0.6.           bus   ~ taxi = 0.4.
\end{verbatim}

\noindent We also state that the t-norm to be used in the transitive closure is the conjunction of G\"{o}del
(i.e., the infimum between two elements).
For this program (the set of rules of Figure \ref{fig-program-goal}, the lattice $L$ and the similarity relation, $\cR$, just described before), the goal {\tt good\_hotel(X)} produces two fuzzy computed answers: {\tt <0.4, \{X/ritz\}>} and
{\tt <0.38, \{X/hydropolis\}>}.
Each one corresponds to the leaves of the tree\footnote{Each state contains its corresponding goal and substitution components and they are drawn inside yellow 
ovals. 
Computational steps, colored in blue, are labeled with the program rule they exploit in the case of {\it successful} steps or 
the annotation ``R0''
in the case of {\it failure} steps (observe that, ``R0'' is a simple notation and do not correspond 
with any existing rule).
Finally,
the blue circles annotated with the word ``is'', correspond to {\it interpretive} steps.}  depicted in Figure \ref{fig-tree}.
Note that for reaching these solutions, a {\it failure} step was performed in the derivation of the left-most  branch, whereas in the right-most one (and this is the crucial novelty w.r.t. previous versions of the \fl\ tool) there exist two  {\it successful} steps exploiting the similarity relation 
atom
{\tt close(hydropolis,metro)}, which illustrates the flexibility of our system.
\end{example}

Ending this section, it is worthy to say that our approach differs from the one presented 
in \cite{CRR14TPLP} since they employ a combination of transformation techniques to first extract
the definition of a predicate ``$\sim$'', simulating weak unification in terms of a set of 
complex program rules that extends the original program.
Finally, this predicate ``$\sim$'' is reduced to a built-in proximity/similarity unification 
operator (in this case not implemented by rules and very close to the implementation of our 
weak unification algorithm) that highly improves the efficiency of their previous programming systems. 

\subsection{FLOPER online}\label{sub-sec-onl}

Aside from the textual and graphical interfaces seen above, we have recently developed a web page aiming at offering the FLOPER system via the Internet, without requiring any further installation.
The interaction with the system is possible through the URL 
\url{http://dectau.uclm.es/floper/?q=sim/test}. Under the title of \emph{FLOPER Online} it shows an interface divided in two areas. 
The \emph{Input} area, shown in Figure~$4$, 
is located in the upper part of the window, and the \emph{Output} area is in the lower part of the window, and is illustrated by Figure~$5$. 

\begin{figure}[t] \vspace{-0.2cm}
   \label{fig-webInput}
   \center
	\includegraphics[width=0.85\textwidth]{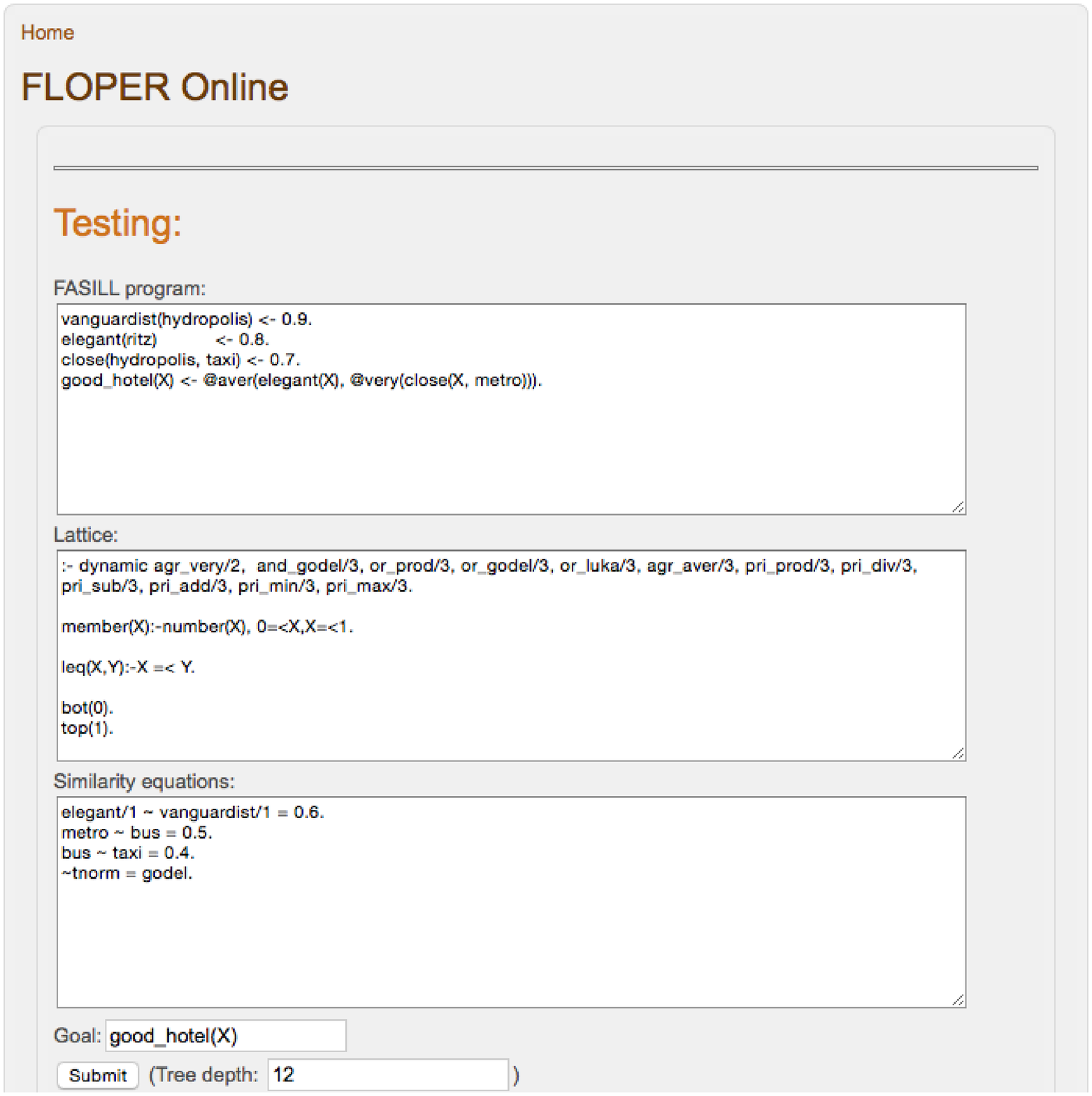}
   \caption{Screenshot of the FLOPER online tool input}\vspace{-0.2cm}
\end{figure}

The \emph{Input} area shows three boxes. The first one, under the label ``FASILL program'' is intended to contain a set of FASILL rules, that is, the fuzzy program. The second one contains the lattice associated to the previously introduced program. By default it includes the $([0,1], \leq)$ lattice, obviously expressed as a \pr\ program following the restrictions previously detailed, with the usual operators. The user is free to implement here any complete lattice as far as it fulfils the syntactic constrains. In the third box the user can write a set of similarity equations using the program's signature. 
After these boxes the user can introduce a goal in a text box (in Figure~$4$, 
the goal is \texttt{good\_hotel(X)}). Finally, by clicking the \emph{Submit} button, the fuzzy program, together with its associated lattice and the similarity relation, is sent to the server with the goal to be executed. 

\begin{figure}[t] \vspace{-0.2cm}
   \label{fig-webOutput}
   \center
	\includegraphics[width=0.85\textwidth]{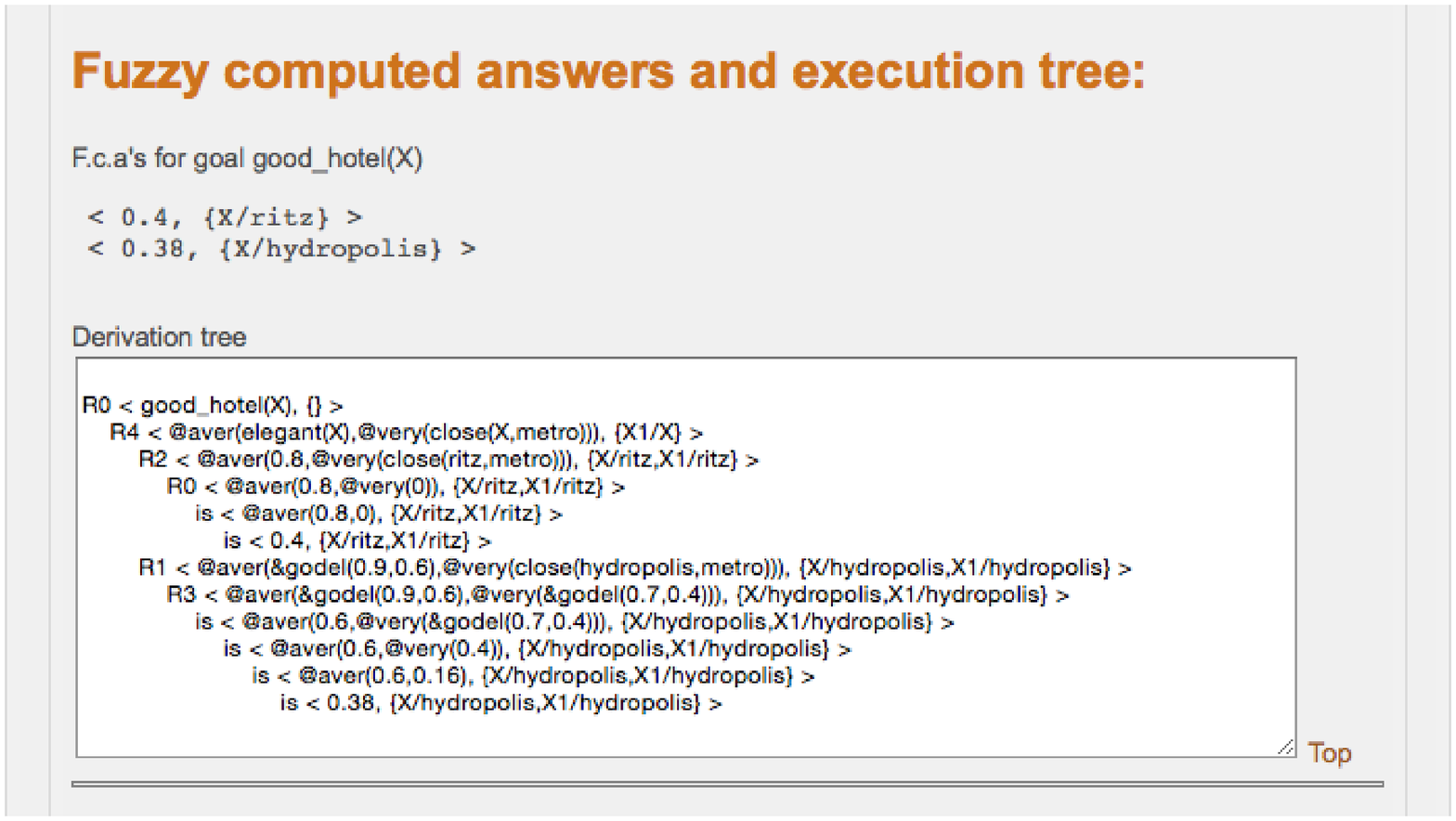}
   \caption{Screenshot of the FLOPER online tool output}
\end{figure}

The result appears in the \emph{Output} area in two ways. In the first place, under the label \emph{F.c.a's for goal \ldots} (including the proper goal), the system shows the fuzzy computed answers for the introduced program and goal. In the figure this corresponds to \texttt{< 0.4, \{X/Ritz\} >} and \texttt{< 0.38, \{X/hydropolis\} >}, as expected. Further, in the box below, the derivation tree is depicted in a textual way.

The tool has been implemented as a php page inside the web of \fa. This php document sends the content of the text boxes (the \fa\ program, the lattice, the similarity equations and the goal) to itself via the ``post'' method. When the php loads again with non empty ``post'' parameters, it creates files in the server to host the \fa\ program, the lattice and the similarity equations, and calls the \pr\ interpreter. Then, it consults the \fl\ environment, loads the files and queries the goal. The output of this task (that is, the corresponding f.c.a's and execution tree) are finally shown in the window.

\section{Conclusions and Future Work}\label{sec-con}

This paper describes an extension of the \fl\ system to cope with the
twofold integrated fuzzy programming language \fa, whose procedural principle is centered upon 
a weak --instead of a syntactic-- unification algorithm based on similarity relations.
After a brief introduction of the syntactic and operational features of \fa, we describe the implementation details of the renewed \fl\ system which gives support to \fa. We center our attention on the description of the {\em similarity module}, providing insights of the internal representation of a similarity relation and its automatic construction, via built-in closure algorithms. Also we describe 
the new tool of {\em FLOPER online}, that allows the execution of \fa\ programs thru the web.

On the other hand, in \cite{JR09PPDP,JMP09IWANN,MPV14JRS} we provided some advances in the design of declarative semantics and/or correctness properties regarding the development of fuzzy logic languages dealing with similarity/proximity relations (Bousi$\sim$Prolog) or highly expressive lattices modeling truth degrees (\ma).
As a matter of future work we want to establish that analogous --but reinforced-- formal properties 
also hold in the language \fa.


\begin{thebibliography}{10}
\providecommand{\bibitemdeclare}[2]{}
\providecommand{\surnamestart}{}
\providecommand{\surnameend}{}
\providecommand{\urlprefix}{Available at }
\providecommand{\url}[1]{\texttt{#1}}
\providecommand{\href}[2]{\texttt{#2}}
\providecommand{\urlalt}[2]{\href{#1}{#2}}
\providecommand{\doi}[1]{doi:\urlalt{http://dx.doi.org/#1}{#1}}
\providecommand{\bibinfo}[2]{#2}

\bibitemdeclare{article}{FA02}
\bibitem{FA02}
\bibinfo{author}{F.~\surnamestart Arcelli\surnameend} (\bibinfo{year}{2002}):
  \emph{\bibinfo{title}{Likelog for flexible query answering}}.
\newblock {\sl \bibinfo{journal}{Soft Computing}}
  \bibinfo{volume}{7}(\bibinfo{number}{2}), pp. \bibinfo{pages}{107--114}.
  \newblock \urlprefix\url{http://dx.doi.org/10.1007/s00500-002-0178-6}.

\bibitemdeclare{inproceedings}{FF99}
\bibitem{FF99}
\bibinfo{author}{F.~\surnamestart Arcelli\surnameend} \&
  \bibinfo{author}{F.~\surnamestart Formato\surnameend} (\bibinfo{year}{1999}):
  \emph{\bibinfo{title}{Likelog: A Logic Programming Language for Flexible Data
  Retrieval}}.
\newblock In: {\sl \bibinfo{booktitle}{Proc. of the 1999 ACM Symposium on
  Applied Computing, SAC'99, San Antonio, Texas}}, pp.
  \bibinfo{pages}{260--267}.
  \newblock \urlprefix\url{http://dx.doi.org/10.1145/298151.298348}.

\bibitemdeclare{article}{FF02}
\bibitem{FF02}
\bibinfo{author}{F.~\surnamestart Arcelli\surnameend} \&
  \bibinfo{author}{F.~\surnamestart Formato\surnameend} (\bibinfo{year}{2002}):
  \emph{\bibinfo{title}{A similarity-based resolution rule.}}
\newblock {\sl \bibinfo{journal}{International Journal of Intelligent Systems}}
  \bibinfo{volume}{17}(\bibinfo{number}{9}), pp. \bibinfo{pages}{853--872}.
  \newblock \urlprefix\url{http://dx.doi.org/10.1002/int.10067}.

\bibitemdeclare{inproceedings}{FFG96}
\bibitem{FFG96}
\bibinfo{author}{F.~\surnamestart Arcelli\surnameend},
  \bibinfo{author}{F.~\surnamestart Formato\surnameend} \&
  \bibinfo{author}{G.~\surnamestart Gerla\surnameend} (\bibinfo{year}{1996}):
  \emph{\bibinfo{title}{Similitude-based unification as a foundation of fuzzy
  logic programming}}.
\newblock In: {\sl \bibinfo{booktitle}{Proc. of Int. Workshop of Logic
  Programming and Soft Computing, Bonn}}.

\bibitemdeclare{inproceedings}{CRR08}
\bibitem{CRR08}
\bibinfo{author}{R.~\surnamestart Caballero\surnameend},
  \bibinfo{author}{M.~\surnamestart Rodr\'{\i}guez-Artalejo\surnameend} \&
  \bibinfo{author}{C.~A. \surnamestart Romero-D\'{\i}az\surnameend}
  (\bibinfo{year}{2008}): \emph{\bibinfo{title}{Similarity-based reasoning in
  qualified logic programming}}.
\newblock In: {\sl \bibinfo{booktitle}{Proc. of the 10th Int. ACM SIGPLAN
  Conference on Principles and Practice of Declarative Programming}},
  \bibinfo{series}{PPDP'08}, \bibinfo{publisher}{ACM}, \bibinfo{address}{New
  York, USA}, pp. \bibinfo{pages}{185--194}.
  \newblock \urlprefix\url{http://dx.doi.org/10.1145/1389449.1389472}.

\bibitemdeclare{article}{CRR14TPLP}
\bibitem{CRR14TPLP}
\bibinfo{author}{R.~\surnamestart Caballero\surnameend},
  \bibinfo{author}{M.~\surnamestart Rodr\'{\i}guez-Artalejo\surnameend} \&
  \bibinfo{author}{C.~A. \surnamestart Romero-D\'{\i}az\surnameend}
  (\bibinfo{year}{2014}): \emph{\bibinfo{title}{A Transformation-based
  implementation for CLP with qualification and proximity}}.
\newblock {\sl \bibinfo{journal}{Theory and Practice of Logic Programming}}
  \bibinfo{volume}{14}(\bibinfo{number}{1}), pp. \bibinfo{pages}{1--63}.
  \newblock \urlprefix\url{http://dx.doi.org/10.1017/S1471068412000014}.

\bibitemdeclare{article}{Emd86}
\bibitem{Emd86}
\bibinfo{author}{M.~H. \surnamestart van Emden\surnameend}
  (\bibinfo{year}{1986}): \emph{\bibinfo{title}{Quantitative Deduction and its
  Fixpoint Theory}}.
\newblock {\sl \bibinfo{journal}{Journal of Logic Programming}}
  \bibinfo{volume}{3}(\bibinfo{number}{1}), pp. \bibinfo{pages}{37--53}.
  \newblock \urlprefix\url{http://dx.doi.org/10.1016/0743-1066(86)90003-8}.

\bibitemdeclare{inproceedings}{FGS99}
\bibitem{FGS99}
\bibinfo{author}{F.~\surnamestart Formato\surnameend},
  \bibinfo{author}{G.~\surnamestart Gerla\surnameend} \& \bibinfo{author}{M.~I.
  \surnamestart Sessa\surnameend} (\bibinfo{year}{1999}):
  \emph{\bibinfo{title}{Extension of Logic Programming by Similarity.}}
\newblock In: {\sl \bibinfo{booktitle}{Proc. of 1999 Joint Conference on
  Declarative Programming, AGP'99, L'Aquila, Italy}}, pp.
  \bibinfo{pages}{397--410}.

\bibitemdeclare{article}{FGS00}
\bibitem{FGS00}
\bibinfo{author}{F.~\surnamestart Formato\surnameend},
  \bibinfo{author}{Giangiacomo \surnamestart Gerla\surnameend} \&
  \bibinfo{author}{Maria~I. \surnamestart Sessa\surnameend}
  (\bibinfo{year}{2000}): \emph{\bibinfo{title}{Similarity-based Unification.}}
\newblock {\sl \bibinfo{journal}{Fundamenta Informaticae}}
  \bibinfo{volume}{41}(\bibinfo{number}{4}), pp. \bibinfo{pages}{393--414}.
  \newblock \urlprefix\url{http://dx.doi.org/10.3233/FI-2000-41402}.

\bibitemdeclare{inproceedings}{JMP09IWANN}
\bibitem{JMP09IWANN}
\bibinfo{author}{P.~\surnamestart Juli{\'a}n\surnameend},
  \bibinfo{author}{G.~\surnamestart Moreno\surnameend} \&
  \bibinfo{author}{J.~\surnamestart Penabad\surnameend} (\bibinfo{year}{2009}):
  \emph{\bibinfo{title}{On the Declarative Semantics of Multi-Adjoint Logic
  Programs}}.
\newblock In: {\sl \bibinfo{booktitle}{Proc. of 10th Int. Work-Conference on
  Artificial Neural Networks (Part I), IWANN'09}}, \bibinfo{publisher}{Lectures
  Notes in Computer Science, 5517, Springer Verlag}, pp.
  \bibinfo{pages}{253--260}.
  \newblock \urlprefix\url{http://dx.doi.org/10.1007/978-3-642-02478-8_32}.
  

\bibitemdeclare{inproceedings}{JR09PPDP}
\bibitem{JR09PPDP}
\bibinfo{author}{P.~\surnamestart Juli{\'a}n-Iranzo\surnameend} \&
  \bibinfo{author}{C.~\surnamestart Rubio-Manzano\surnameend}
  (\bibinfo{year}{2009}): \emph{\bibinfo{title}{A declarative semantics for
  Bousi$\sim$Prolog}}.
\newblock In: {\sl \bibinfo{booktitle}{Proc. of 11th Int. ACM SIGPLAN
  Conference on Principles and Practice of Declarative Programming, PPDP'09,
  Coimbra, Portugal}}, \bibinfo{publisher}{ACM}, pp. \bibinfo{pages}{149--160}.
  \newblock \urlprefix\url{http://doi.acm.org/10.1145/1599410.1599430}.

\bibitemdeclare{inproceedings}{JR09IWANN}
\bibitem{JR09IWANN}
\bibinfo{author}{P.~\surnamestart Juli{\'a}n-Iranzo\surnameend} \&
  \bibinfo{author}{C.~\surnamestart Rubio-Manzano\surnameend}
  (\bibinfo{year}{2009}): \emph{\bibinfo{title}{A Similarity-Based WAM for
  {B}ousi$\sim${P}rolog}}.
\newblock In \bibinfo{editor}{J.~Cabestany \surnamestart et~al.\surnameend},
  editor: {\sl \bibinfo{booktitle}{Proc. of 10th Int. Work-Conference on
  Artificial Neural Networks, IWANN'09, Part I, Salamanca, Spain, 2009}}, {\sl
  \bibinfo{series}{Lecture Notes in Computer Science}} \bibinfo{volume}{5517},
  \bibinfo{publisher}{Springer}, pp. \bibinfo{pages}{245--252}.
  \newblock \urlprefix\url{http://dx.doi.org/10.1007/978-3-642-02478-8_31}.

\bibitemdeclare{inproceedings}{JR10FUZZIEEE}
\bibitem{JR10FUZZIEEE}
\bibinfo{author}{P.~\surnamestart Juli{\'a}n-Iranzo\surnameend} \&
  \bibinfo{author}{C.~\surnamestart Rubio-Manzano\surnameend}
  (\bibinfo{year}{2010}): \emph{\bibinfo{title}{An efficient fuzzy unification
  method and its implementation into the Bousi$\sim$Prolog system}}.
\newblock In: {\sl \bibinfo{booktitle}{Proc. of the 2010 IEEE Int. Conference
  on Fuzzy Systems}}, pp. \bibinfo{pages}{1--8}.
  \newblock \urlprefix\url{http://dx.doi.org/10.1109/FUZZY.2010.5584193}.

\bibitemdeclare{inproceedings}{JR11IWANN}
\bibitem{JR11IWANN}
\bibinfo{author}{P.~\surnamestart Juli{\'a}n-Iranzo\surnameend} \&
  \bibinfo{author}{C.~\surnamestart Rubio-Manzano\surnameend}
  (\bibinfo{year}{2011}): \emph{\bibinfo{title}{{A Sound Semantics for a
  Similarity-Based Logic Programming Language}}}.
\newblock In: {\sl \bibinfo{booktitle}{Proc. of 11th Int. Work-Conference on
  Artificial Neural Networks, IWANN'11, Part II, Torremolinos, M{\'a}laga,
  Spain, 2011}}, {\sl \bibinfo{series}{Lecture Notes in Computer Science}}
  \bibinfo{volume}{6692}, \bibinfo{publisher}{Springer}, pp.
  \bibinfo{pages}{421--428}.
\newblock \urlprefix\url{http://dx.doi.org/10.1007/978-3-642-21498-1_53}.

\bibitemdeclare{article}{JRG09ENTCS}
\bibitem{JRG09ENTCS}
\bibinfo{author}{P.~\surnamestart Juli{\'a}n-Iranzo\surnameend},
  \bibinfo{author}{C.~\surnamestart Rubio-Manzano\surnameend} \&
  \bibinfo{author}{J.~\surnamestart Gallardo-Casero\surnameend}
  (\bibinfo{year}{2009}): \emph{\bibinfo{title}{Bousi$\sim$Prolog: a Prolog
  Extension Language for Flexible Query Answering}}.
\newblock {\sl \bibinfo{journal}{Electronic Notes in Theoretical Computer
  Science}} \bibinfo{volume}{248}, pp. \bibinfo{pages}{131--147}.
  \newblock \urlprefix\url{http://dx.doi.org/10.1016/j.entcs.2009.07.064}.

\bibitemdeclare{article}{KS92}
\bibitem{KS92}
\bibinfo{author}{M.~\surnamestart Kifer\surnameend} \& \bibinfo{author}{V.S.
  \surnamestart Subrahmanian\surnameend} (\bibinfo{year}{1992}):
  \emph{\bibinfo{title}{Theory of generalized annotated logic programming and
  its applications.}}
\newblock {\sl \bibinfo{journal}{Journal of Logic Programming}}
  \bibinfo{volume}{12}, pp. \bibinfo{pages}{335--367}.
  \newblock \urlprefix\url{http://dx.doi.org/10.1016/0743-1066(92)90007-P}.

\bibitemdeclare{inproceedings}{LSS01}
\bibitem{LSS01}
\bibinfo{author}{V.~\surnamestart Loia\surnameend},
  \bibinfo{author}{S.~\surnamestart Senatore\surnameend} \&
  \bibinfo{author}{M.~I. \surnamestart Sessa\surnameend}
  (\bibinfo{year}{2001}): \emph{\bibinfo{title}{Similarity-based SLD Resolution
  and Its Implementation in An Extended Prolog System}}.
\newblock In: {\sl \bibinfo{booktitle}{Proc. of 10th IEEE Int. Conference on
  Fuzzy Systems, FUZZ-IEEE'01, Melbourne, Australia}}, pp.
  \bibinfo{pages}{650--653}.
  \newblock \urlprefix\url{http://dx.doi.org/10.1109/FUZZ.2001.1009038}.

\bibitemdeclare{article}{MM82}
\bibitem{MM82}
\bibinfo{author}{A.~\surnamestart Martelli\surnameend} \&
  \bibinfo{author}{U.~\surnamestart Montanari\surnameend}
  (\bibinfo{year}{1982}): \emph{\bibinfo{title}{An {E}fficient {U}nification
  {A}lgorithm}}.
\newblock {\sl \bibinfo{journal}{ACM Transactions on Programming Languages and
  Systems}} \bibinfo{volume}{4}, pp. \bibinfo{pages}{258--282}.
  \newblock \urlprefix\url{http://dx.doi.org/10.1145/357162.357169}.

\bibitemdeclare{article}{MOV04}
\bibitem{MOV04}
\bibinfo{author}{J.~\surnamestart Medina\surnameend},
  \bibinfo{author}{M.~\surnamestart Ojeda-Aciego\surnameend} \&
  \bibinfo{author}{P.~\surnamestart Vojt\'a\v{s}\surnameend}
  (\bibinfo{year}{2004}): \emph{\bibinfo{title}{Similarity-based {U}nification:
  a multi-adjoint approach}}.
\newblock {\sl \bibinfo{journal}{{F}uzzy {S}ets and {S}ystems}}
  \bibinfo{volume}{146}, pp. \bibinfo{pages}{43--62}.
  \newblock \urlprefix\url{http://dx.doi.org/10.1016/j.fss.2003.11.005}.

\bibitemdeclare{inproceedings}{MMPV10RULEML}
\bibitem{MMPV10RULEML}
\bibinfo{author}{P.~J. \surnamestart Morcillo\surnameend},
  \bibinfo{author}{G.~\surnamestart Moreno\surnameend},
  \bibinfo{author}{J.~\surnamestart Penabad\surnameend} \&
  \bibinfo{author}{C.~\surnamestart V{\'a}zquez\surnameend}
  (\bibinfo{year}{2010}): \emph{\bibinfo{title}{{A} {P}ractical {M}anagement of
  {F}uzzy {T}ruth {D}egrees using {FLOPER}}}.
\newblock In: {\sl \bibinfo{booktitle}{Proc. of 4nd Int. Symposium on Rule
  Interchange and Applications, RuleML'10}}, \bibinfo{publisher}{Lectures Notes
  in Computer Science, 6403, Springer Verlag}, pp. \bibinfo{pages}{20--34}.
  \newblock \urlprefix\url{http://dx.doi.org/10.1007/978-3-642-16289-3_4}.

\bibitemdeclare{inproceedings}{MPV13CMMSE}
\bibitem{MPV13CMMSE}
\bibinfo{author}{G.~\surnamestart Moreno\surnameend},
  \bibinfo{author}{J.~\surnamestart Penabad\surnameend} \&
  \bibinfo{author}{C.~\surnamestart V{\'a}zquez\surnameend}
  (\bibinfo{year}{2013}): \emph{\bibinfo{title}{Relaxing the Role of Adjoint
  Pairs in Multi-adjoint Logic Programming}}.
\newblock In \bibinfo{editor}{I.~\surnamestart Hamilton\surnameend} \&
  \bibinfo{editor}{J.~\surnamestart Vigo-Aguiar\surnameend}, editors: {\sl
  \bibinfo{booktitle}{Proc. of 13th Int. Conference on Mathematical Methods in
  Science and Engineering, CMMSE'13 (Volume III), Cabo de Gata, Almer\'{\i}a}},
  pp. \bibinfo{pages}{1156--1167}.
\newblock \urlprefix\url{http://cmmse.usal.es/images/stories/congreso/volume3-cmmse-20013.pdf}.

\bibitemdeclare{inproceedings}{MPV14JRS}
\bibitem{MPV14JRS}
\bibinfo{author}{G.~\surnamestart Moreno\surnameend},
  \bibinfo{author}{J.~\surnamestart Penabad\surnameend} \&
  \bibinfo{author}{C.~\surnamestart V{\'a}zquez\surnameend}
  (\bibinfo{year}{2014}): \emph{\bibinfo{title}{Fuzzy Sets for a Declarative
  Description of Multi-adjoint Logic Programming}}.
\newblock In: {\sl \bibinfo{booktitle}{Proc. of the 2014 Joint Rough Set
  Symposium, JRS'14}}, \bibinfo{publisher}{Lecture Notes in Computer Science,
  8536, Springer Verlag}, pp. \bibinfo{pages}{71--82}.
\newblock \urlprefix\url{http://dx.doi.org/10.1007/978-3-319-08644-6_7}.


\bibitemdeclare{article}{MV14JSEA}
\bibitem{MV14JSEA}
\bibinfo{author}{G.~\surnamestart Moreno\surnameend} \&
  \bibinfo{author}{C.~\surnamestart V\'azquez\surnameend}
  (\bibinfo{year}{2014}): \emph{\bibinfo{title}{Fuzzy Logic Programming in
  Action with FLOPER}}.
\newblock {\sl \bibinfo{journal}{Journal of Software Engineering and
  Applications}} \bibinfo{volume}{7}, pp. \bibinfo{pages}{237--298}.
\newblock \urlprefix\url{http://dx.doi.org/10.4236/jsea.2014.74028}.

\bibitemdeclare{article}{MPH11INS}
\bibitem{MPH11INS}
\bibinfo{author}{S.~\surnamestart Mu{\~n}oz-Hern{\'a}ndez\surnameend},
  \bibinfo{author}{V.~P. \surnamestart Ceruelo\surnameend} \&
  \bibinfo{author}{H.~\surnamestart Strass\surnameend} (\bibinfo{year}{2011}):
  \emph{\bibinfo{title}{RFuzzy: Syntax, semantics and implementation details of
  a simple and expressive fuzzy tool over Prolog}}.
\newblock {\sl \bibinfo{journal}{Information Sciences}}
  \bibinfo{volume}{181}(\bibinfo{number}{10}), pp. \bibinfo{pages}{1951--1970}.
  \newblock \urlprefix\url{http://dx.doi.org/10.1016/j.ins.2010.07.033}.

\bibitemdeclare{book}{NW06}
\bibitem{NW06}
\bibinfo{author}{H.~T. \surnamestart Nguyen\surnameend} \&
  \bibinfo{author}{E.~A. \surnamestart Walker\surnameend}
  (\bibinfo{year}{2006}): \emph{\bibinfo{title}{A First Course in Fuzzy
  Logic}}.
\newblock \bibinfo{publisher}{Chatman \& Hall}, \bibinfo{address}{Boca Rat\'on,
  Florida}.
  \newblock \urlprefix\url{http://www.crcpress.com/product/isbn/9780849316593}.

\bibitemdeclare{inproceedings}{RR08FLOPS}
\bibitem{RR08FLOPS}
\bibinfo{author}{M.~\surnamestart Rodr\'{\i}guez-Artalejo\surnameend} \&
  \bibinfo{author}{C.~\surnamestart Romero-D\'{\i}az\surnameend}
  (\bibinfo{year}{2008}): \emph{\bibinfo{title}{Quantitative logic programming
  revisited}}.
\newblock In \bibinfo{editor}{J.~\surnamestart Garrigue\surnameend} \&
  \bibinfo{editor}{M.~\surnamestart Hermenegildo\surnameend}, editors: {\sl
  \bibinfo{booktitle}{Proc. of 9th Functional and Logic Programming Symposium,
  FLOPS'08}}, \bibinfo{publisher}{Lecture Notes in Computer Science 4989,
  Springer}, pp. \bibinfo{pages}{272--288}.
  \newblock \urlprefix\url{http://dx.doi.org/10.1007/978-3-540-78969-7_20}.

\bibitemdeclare{article}{RR10}
\bibitem{RR10}
\bibinfo{author}{M.~\surnamestart Rodr\'{\i}guez-Artalejo\surnameend} \&
  \bibinfo{author}{C.~A. \surnamestart Romero-D\'{\i}az\surnameend}
  (\bibinfo{year}{2008}): \emph{\bibinfo{title}{A declarative semantics for clp
  with qualification and proximity}}.
\newblock {\sl \bibinfo{journal}{Theory and Practice of Logic Programing}}
  \bibinfo{volume}{10}, pp. \bibinfo{pages}{627--642}.
  \newblock \urlprefix\url{http://dx.doi.org/10.1017/S1471068410000323}.

\bibitemdeclare{article}{RJ14JIFS}
\bibitem{RJ14JIFS}
\bibinfo{author}{C.~\surnamestart Rubio-Manzano\surnameend} \&
  \bibinfo{author}{P.~\surnamestart Juli{\'a}n-Iranzo\surnameend}
  (\bibinfo{year}{2014}): \emph{\bibinfo{title}{A Fuzzy linguistic prolog and
  its applications}}.
\newblock {\sl \bibinfo{journal}{Journal of Intelligent and Fuzzy Systems}}
  \bibinfo{volume}{26}(\bibinfo{number}{3}), pp. \bibinfo{pages}{1503--1516}.
\newblock \urlprefix\url{http://dx.doi.org/10.3233/IFS-130834}.

\bibitemdeclare{article}{Ses02}
\bibitem{Ses02}
\bibinfo{author}{M.~I. \surnamestart Sessa\surnameend} (\bibinfo{year}{2002}):
  \emph{\bibinfo{title}{Approximate reasoning by similarity-based SLD
  resolution.}}
\newblock {\sl \bibinfo{journal}{Theoretical Computer Science}}
  \bibinfo{volume}{275}(\bibinfo{number}{1-2}), pp. \bibinfo{pages}{389--426}.
\newblock \urlprefix\url{http://dx.doi.org/10.1016/S0304-3975(01)00188-8}.

\end{thebibliography}
\end{document}